\documentclass[12pt,a4paper,english]{article}
\usepackage{float}
\usepackage{setspace}
\makeatletter

\providecommand{\tabularnewline}{\\}

\makeatother

\usepackage{babel}

\begin{document}
\begin{doublespace}

\title{Quantum Teleportation with Remote Rotation on a GHZ state}
\end{doublespace}

\author{Jung-Lun Hsu, Yu-Ting Chen, Chia-Wei Tsai and Tzonelish Hwang%
\thanks{Corresponding Author%
}}

\date{$\;$}
\maketitle
\begin{abstract}
This study proposes a pioneering protocol for teleporting an arbitrary
single particle state and simultaneously performing a rotation operation
on that particle. There are protocols for either only teleporting
particles or only remotely controlling quantum particles. If one has
to remotely control a teleported quantum, then he/she has to first
do the quantum teleportation and then perform the remote control on
the teleported quantum. Both operations were done separately on two
sets of entanglements. However, this intuitive solution is inefficient
because many resources are wasted. Therefore, the study attempts to
complete both operations using only one Greenberger-Horne-Zeilinger
(GHZ) state. 
\end{abstract}

\section{Introduction}

The study of quantum information theory is an extended research in
recent years. In the quantum information theory, the entanglement
is a key physical property and has been employed in many applications
(e.g., quantum teleportation \cite{CQT_ES_Zhang_2005,CQT_GHZ_Karlsson_1998,CQT_Li_2007,CQT_Wang_2011,CQT_Yan_2003,CQT_Yang_2004,QT_EPR_Bennett_1993,QT_GHZ-like_Banerjee_2011,QT_GHZ-like_Yang_2009,QT_Gorbachev_2000,QT_Joo_2003,QT_Li_2007,QT_Pankaj_2006,QT_Tsai_2010},
quantum key distribution \cite{AQKD_Li_2007,MQKD_Tsai_2010,QKD_Hwang_2011,QKD_Lee_2007,QKD_QA_Wei_2011,QKD_Shih_2009},
quantum secret sharing \cite{MQSS_Hwang_2011,MQSS_Li_2006,MQSS_Lin_2011,MQSS_Shi_2010,MQSS_Wang_2010,MQSS_Yan_2005,QSS_Chen_2009,QSS_Gottesman_2000,QSS_Hsieh_2010,QSS_QECC_Zhang_2011}
and etc.). Among these applications, \emph{quantum teleportation}
is an epochal application, in which an arbitrary particle can be transmitted
to a distant place through the property of entanglement and some auxiliary
classical communications without using any physical quantum channel.
In 1993, Bennett and Brassard proposed the first teleportation protocol
\cite{QT_EPR_Bennett_1993}, in which two participants (called as
Alice and Bob) in a distance away pre-share a two-particle entanglement
quantum state, Einstein-Podolsky-Rosen (EPR) state. And then Alice
can use the entanglement of EPR state to teleport an arbitrary single
particle ($\alpha\left|0\right\rangle +\beta\left|1\right\rangle $)
to Bob. Since Bennett and Brassard's teleportation protocol \cite{QT_EPR_Bennett_1993},
many quantum teleportation protocols \cite{CQT_ES_Zhang_2005,CQT_GHZ_Karlsson_1998,CQT_Li_2007,CQT_Wang_2011,CQT_Yan_2003,CQT_Yang_2004,QT_GHZ-like_Banerjee_2011,QT_GHZ-like_Yang_2009,QT_Gorbachev_2000,QT_Joo_2003,QT_Li_2007,QT_Pankaj_2006,QT_Tsai_2010}
have been introduced to teleport various qubits via different entanglement
quantum state (e.g., GHZ class state, W class state, and etc.) subsequently. 

Except for teleporting an arbitrary particle, the entanglement can
also be employed to transmit the information of a unitary operator.
Suppose that a participant, Alice, wants to perform an arbitrary unitary
operator on a particle held by a remote participant, Bob, but she
is unable to immediately transmit the information of a unitary operator
to Bob or she would not like to let Bob know what the operator is.
Similar to quantum teleportation, Alice and Bob can use the correlation
of a pre-shared entanglement state to complete the task. This is the
so called \emph{quantum remote control}. Since the first quantum remote
control protocol presented by Huelga et al. \cite{QRC_Huelga_2001}
in 2001, various quantum remote control protocols \cite{QRC_Chen_2005,QRC_Chen_2007,QRC_Huelga_2002,QRC_Yao_2006,QRC_Zhang_2004}
have been introduced. 

Let us consider a situation which may be useful in quantum cryptography,
such as quantum secret sharing or quantum private comparison and etc..
Bob wants to teleport an arbitrary qubit $\left|\varphi\right\rangle =\alpha\left|0\right\rangle +\beta\left|1\right\rangle $
to Charlie, and Alice wants to do a unitary operator on the particle
$\left|\varphi\right\rangle $. In order to achieve the task by using
the existing protocols, an intuitive solution can be adopted. That
is, Bob first uses the quantum teleportation with an EPR state to
teleport the particle $\left|\varphi\right\rangle $ to Charlie, and
then Alice employs the quantum remote control with the other EPR state
to perform her operation on the particle $\left|\varphi\right\rangle $
held by Charlie. In total, 2 EPR states, 5 unitary operators and 4
classical bits were needed in this intuitive solution. This paper
attempts to provide a more efficient protocol for this situation.
We aim to perform the quantum teleportation and quantum remote control
simultaneously by using one three-particle GHZ state. 

The rest of this paper is constructed as follows. Section 2 first
introduces the proposed protocol. Then, Section 3 shows a comparison
of our protocol to the intuitive solution. Finally, a short conclusion
is given in Section 4.

\section{The Proposed Protocol}

Assume there are three participants Alice, Bob and Charlie in the
protocol. Suppose that Bob wants to teleport an arbitrary single particle
$\left|\psi\right\rangle _{b}=\alpha\left|0\right\rangle _{b}+\beta\left|1\right\rangle _{b}$
to Charlie. At the same time, Alice wants to remotely perform a rotation
operator $R_{z}\left(\theta\right)$ (about the Z axis) on $\left|\psi\right\rangle _{b}$,
where $\theta\in\left[0,\:2\pi\right]$,

\begin{equation}
R_{z}\left(\theta\right)\equiv e^{-i\theta Z/2}=\cos\frac{\theta}{2}I-i\sin\frac{\theta}{2}Z=\left[\begin{array}{cc}
e^{-i\theta/2} & 0\\
0 & e^{i\theta/2}\end{array}\right].\end{equation}
This will produce a new quantum state: \begin{equation}
\left|\varphi\right\rangle _{b}=R_{z}\left(\theta\right)\left|\psi\right\rangle _{b}=\alpha e^{-i\theta/2}\left|0\right\rangle _{b}+\beta e^{i\theta/2}\left|1\right\rangle _{b}.\label{eq:3-1}\end{equation}
 In order to complete the above tasks, let Alice, Bob and Charlie
pre-share a three-particle GHZ state \begin{equation}
\left|\phi\right\rangle _{q_{1}q_{2}q_{3}}=\frac{1}{\sqrt{2}}\left(\left|000\right\rangle +\left|111\right\rangle \right)_{q_{1}q_{2}q_{3}},\end{equation}
where Alice, Bob and Charlie have the first, the second and the third
particles, respectively. The proposed protocol is described in the
following steps:
\begin{description}
\item [{Step1}] Bob first performs a joint measurement on his particle
pair (\emph{b}, $q_{2}$) with the Bell basis $\left\{ \left|\Phi^{\pm}\right\rangle =\left(\left|00\right\rangle \pm\left|11\right\rangle \right)/\sqrt{2},\left|\Psi^{\pm}\right\rangle =\left(\left|01\right\rangle \pm\left|10\right\rangle \right)/\sqrt{2}\right\} $.
The state of the composite quantum system is showed as follow\begin{eqnarray}
\left|\Pi\right\rangle _{bq_{1}q_{2}q_{3}} & = & \left|\psi\right\rangle _{b}\left|\phi\right\rangle _{q_{1}q_{2}q_{3}}\nonumber \\
 & = & \frac{1}{2}[\left|\Phi^{+}\right\rangle _{bq_{2}}\left(\alpha\left|00\right\rangle +\beta\left|11\right\rangle \right)_{q_{1}q_{3}}\nonumber \\
 &  & +\left|\Phi^{-}\right\rangle _{bq_{2}}\left(\alpha\left|00\right\rangle -\beta\left|11\right\rangle \right)_{q_{1}q_{3}}\nonumber \\
 &  & +\left|\Psi^{+}\right\rangle _{bq_{2}}\left(\alpha\left|11\right\rangle +\beta\left|00\right\rangle \right)_{q_{1}q_{3}}\nonumber \\
 &  & +\left|\Psi^{-}\right\rangle _{bq_{2}}\left(\alpha\left|11\right\rangle -\beta\left|00\right\rangle \right)_{q_{1}q_{3}}].\end{eqnarray}
After Bob performs the measurement, he broadcasts his measurement
result, $MR_{B}$, to Alice and Charlie via a classical channel.
\item [{Step2}] Alice performs the rotation of an angle $\theta$ on the
first particle of the GHZ state, $q_{1}$, according to $MR_{B}$.
If $MR_{B}$ is $\left|\Phi^{\pm}\right\rangle $, Alice performs
$R_{z}\left(\theta\right)$ on $q_{1}$. Otherwise, Alice performs
$R_{z}\left(-\theta\right)$ on $q_{1}$. Then, Alice measures it
with X basis $\left\{ \left|+\right\rangle =\left(\left|0\right\rangle +\left|1\right\rangle \right)/\sqrt{2},\left|-\right\rangle =\left(\left|0\right\rangle -\left|1\right\rangle \right)/\sqrt{2}\right\} $
to obtain the measurement result, $MR_{A}$, which will be sent to
Charlie via a classical channel. 
\item [{Step3}] According to $MR_{A}$ and $MR_{B}$, Charlie can perform
a corresponding unitary operation (shown in Table \ref{tab:1}) on
$q_{3}$ to adjust the state to the one given in Eq.(2).
\end{description}
\begin{table}[H]
\caption{Measurement results and the corresponding operations. \label{tab:1}}

$\vphantom{}$

\begin{centering}
\begin{tabular}{|c|c|c|}
\hline 
$MR_{A}$ & $MR_{B}$ & operation\tabularnewline
\hline
\hline 
$\left|+\right\rangle $ & $\left|\Phi^{+}\right\rangle $ & $I$\tabularnewline
\hline 
$\left|-\right\rangle $ & $\left|\Phi^{+}\right\rangle $ & $\sigma_{z}$\tabularnewline
\hline 
$\left|+\right\rangle $ & $\left|\Phi^{-}\right\rangle $ & $\sigma_{z}$\tabularnewline
\hline 
$\left|-\right\rangle $ & $\left|\Phi^{-}\right\rangle $ & $I$\tabularnewline
\hline 
$\left|+\right\rangle $ & $\left|\Psi^{+}\right\rangle $ & $\sigma_{x}$\tabularnewline
\hline 
$\left|-\right\rangle $ & $\left|\Psi^{+}\right\rangle $ & $i\sigma_{y}$\tabularnewline
\hline 
$\left|+\right\rangle $ & $\left|\Psi^{-}\right\rangle $ & $i\sigma_{y}$\tabularnewline
\hline 
$\left|-\right\rangle $ & $\left|\Psi^{-}\right\rangle $ & $\sigma_{x}$\tabularnewline
\hline
\end{tabular}
\par\end{centering}

\end{table}

As an example, let $MR_{B}$ be $\left|\Phi^{-}\right\rangle _{bq_{2}}$.
The state of the composite quantum system is showed as follows:\begin{equation}
_{bq_{2}}\left\langle \Phi^{-}\right|\left|\Pi\right\rangle _{bq_{1}q_{2}q_{3}}=\alpha\left|00\right\rangle _{q_{1}q_{3}}-\beta\left|11\right\rangle _{q_{1}q_{3}}.\end{equation}
Suppose Alice's measurement result is $\left|+\right\rangle _{q_{1}}$.
The state of $q_{3}$ held by Charlie will be $\alpha e^{-i\theta/2}\left|0\right\rangle _{q_{3}}-\beta e^{i\theta/2}\left|1\right\rangle _{q_{3}}$
. Once Charlie performs a unitary operation $\sigma_{z}$ on $q_{3}$,
he will obtain the state in Eq.(2).

\section{Comparison}

In the intuitive solution, Bob first has to use the quantum teleportation
to teleport a particle $\left|\psi\right\rangle _{b}$ to Charlie
which requires an EPR state, a corresponding operator and 2 bits of
classical message \cite{QT_EPR_Bennett_1993}. After that, Alice employs
the quantum remote control to perform her operation on the particle
$\left|\psi\right\rangle _{b}$ held by Charlie.

In the remote control protocol, Alice and Charlie also have to pre-share
a Bell state, where the subscripts $q_{1}'$ and $q_{2}'$ represent
the first and the second particles of the Bell state belonged to Alice
and Charlie, respectively,

\begin{equation}
\left|\Phi^{+}\right\rangle _{q_{1}'q_{2}'}=\frac{1}{\sqrt{2}}\left(\left|00\right\rangle +\left|11\right\rangle \right)_{q_{1}'q_{2}'}.\end{equation}

Charlie first performs a Controlled-NOT operation on his qubit pair
($q_{2}'$, $b$) with $q_{2}'$ as the control bit and $b$ as the
target bit. Then, Charlie measures $b$ with Z basis, and the composite
state becomes

\begin{equation}
\left|\omega\right\rangle _{q_{1}'q_{2}'b}=\frac{1}{2}\left[\left(\alpha\left|00\right\rangle +\beta\left|11\right\rangle \right)_{q_{1}'q_{2}'}\left|0\right\rangle _{b}+\left(\alpha\left|11\right\rangle +\beta\left|00\right\rangle \right)_{q_{1}'q_{2}'}\left|1\right\rangle _{b}\right].\end{equation}

Then, Charlie sends a one-bit classical message to Alice about his
measurement. If the measurement result is $\left|0\right\rangle $,
Charlie and Alice will do nothing. Otherwise, they both perform $\sigma_{x}$
on their particles. So the state they shared becomes

\begin{equation}
\left|\omega'\right\rangle _{q_{1}'q_{2}'}=\left(\alpha\left|00\right\rangle +\beta\left|11\right\rangle \right)_{q_{1}'q_{2}'}.\end{equation}

After that, Alice performs $R_{z}\left(\theta\right)$ on her particle
$q_{1}'$ and measures it with X basis. Alice sends a one-bit classical
message to Charlie about her measurement. According to that, Charlie
performs a corresponding unitary operation to complete the remote
control process. Hence, the remote control requires an EPR state,
4 unitary operators and 2 bits of classical message in 2 rounds of
transmission.

On the other hand, in the proposed protocol, a GHZ state is used to
teleport the particle as well as remotely control the particle. In
total, the proposed protocol requires a three-particle GHZ state,
2 unitary operators and 3 bits of classical message in 2 rounds of
transmission. Obviously, it is more efficient than the intuitive solution
(see also Table 2). 

\begin{table}[H]
\caption{The comparison of the proposed protocol to the intuitive solution}
$ $

\centering{}\begin{tabular}{|c|c|c|}
\hline 
 & Intuitive solution & Our protocol\tabularnewline
\hline
\hline 
Entanglement state & 2 EPR states & 1 GHZ state\tabularnewline
\hline 
Unitary operator & 5 & 2\tabularnewline
\hline 
\begin{tabular}{c}
Number of rounds in \tabularnewline
classical transmission\tabularnewline
\end{tabular} & 3 & 2\tabularnewline
\hline 
Classical message & 4 bits & 3 bits\tabularnewline
\hline 
Measurement & \begin{tabular}{c}
1 Bell measurement\tabularnewline
2 single-photon measurement\tabularnewline
\end{tabular}  & \begin{tabular}{c}
1 Bell measurement\tabularnewline
1 single-photon measurement\tabularnewline
\end{tabular}\tabularnewline
\hline
\end{tabular}
\end{table}

\section{Conclusion}

This paper uses the entanglement in GHZ state to perform both the
teleportation of an arbitrary single quantum state and the remote
control of that quantum state simultaneously. The proposed protocol
is more efficient than an intuitive solution, that performs quantum
teleportation and quantum remote control separately in two distinct
entanglements of quantum state. It should be noted that in the proposed
protocol, Alice can only perform the rotation operator under Z axis,
and Bob can only teleport a single particle. How to design a protocol
for Alice to perform an arbitrary operator and Bob to teleport various
quantum states is a promising future research.

\section*{Acknowledgments}

\begin{doublespace}
The authors would like to thank the National Science Council of the
Republic of China, Taiwan for financially supporting this research
under Contract No. NSC 100-2221-E-006-152-MY3.
\end{doublespace}

\bibliographystyle{ieeetr}
\addcontentsline{toc}{section}{\refname}\bibliography{Lab,QT}

\begin{thebibliography}{10}

\bibitem{CQT_ES_Zhang_2005}
Z.-J. Zhang, Y.-M. Liu, and Z.-X. Man, ``Many-agent controlled teleportation of
  multi-qubit quantum information via quantum entanglement swapping,'' {\em
  Communications in Theoretical Physics}, vol.~44, no.~5, p.~847, 2005.

\bibitem{CQT_GHZ_Karlsson_1998}
A.~Karlsson and M.~Bourennane, ``Quantum teleportation using three-particle
  entanglement,'' {\em Phys. Rev. A}, vol.~58, pp.~4394--4400, Dec 1998.

\bibitem{CQT_Li_2007}
X.-H. Li, F.-G. Deng, and H.-Y. Zhou, ``Controlled teleportation of an
  arbitrary multi-qudit state in a general form with d -dimensional
  greenberger–horne–zeilinger states,'' {\em Chinese Physics Letters},
  vol.~24, no.~5, p.~1151, 2007.

\bibitem{CQT_Wang_2011}
T.-Y. Wang and Q.-Y. Wen, ``Controlled quantum teleportation with bell
  states,'' {\em Chinese Physics B}, vol.~20, no.~4, p.~040307, 2011.

\bibitem{CQT_Yan_2003}
F.~Yan and D.~Wang, ``Probabilistic and controlled teleportation of unknown
  quantum states,'' {\em Physics Letters A}, vol.~316, no.~5, pp.~297 -- 303,
  2003.

\bibitem{CQT_Yang_2004}
C.-P. Yang, S.-I. Chu, and S.~Han, ``Efficient many-party controlled
  teleportation of multiqubit quantum information via entanglement,'' {\em
  Phys. Rev. A}, vol.~70, p.~022329, Aug 2004.

\bibitem{QT_EPR_Bennett_1993}
C.~H. Bennett, G.~Brassard, C.~Cr\'epeau, R.~Jozsa, A.~Peres, and W.~K.
  Wootters, ``Teleporting an unknown quantum state via dual classical and
  einstein-podolsky-rosen channels,'' {\em Phys. Rev. Lett.}, vol.~70,
  pp.~1895--1899, Mar 1993.

\bibitem{QT_GHZ-like_Banerjee_2011}
A.~Banerjee, K.~Patel, and A.~Pathak, ``Comment on "quantum teleportation via
  ghz-like state",'' {\em International Journal of Theoretical Physics},
  vol.~50, pp.~507--510, 2011.
\newblock 10.1007/s10773-010-0561-5.

\bibitem{QT_GHZ-like_Yang_2009}
K.~Yang, L.~Huang, W.~Yang, and F.~Song, ``Quantum teleportation via ghz-like
  state,'' {\em International Journal of Theoretical Physics}, vol.~48,
  pp.~516--521, 2009.
\newblock 10.1007/s10773-008-9827-6.

\bibitem{QT_Gorbachev_2000}
V.~N. {Gorbachev}, A.~I. {Trubilko}, A.~I. {Zhiliba}, and E.~S. {Yakovleva},
  ``{Teleportation of entangled states and dense coding using a multiparticle
  quantum channel},'' {\em ArXiv Quantum Physics e-prints}, Nov. 2000.

\bibitem{QT_Joo_2003}
J.~Joo, Y.-J. Park, S.~Oh, and J.~Kim, ``Quantum teleportation via a w state,''
  {\em New Journal of Physics}, vol.~5, no.~1, p.~136, 2003.

\bibitem{QT_Li_2007}
D.-C. Li and Z.-L. Cao, ``Teleportation of two-particle entangled state via
  cluster state,'' {\em Communications in Theoretical Physics}, vol.~47, no.~3,
  p.~464, 2007.

\bibitem{QT_Pankaj_2006}
P.~Agrawal and A.~Pati, ``Perfect teleportation and superdense coding with w
  states,'' {\em Phys. Rev. A}, vol.~74, p.~062320, Dec 2006.

\bibitem{QT_Tsai_2010}
C.-W. Tsai and T.~Hwang, ``Teleportation of a pure epr state via ghz-like
  state,'' {\em International Journal of Theoretical Physics}, vol.~49,
  pp.~1969--1975, 2010.
\newblock 10.1007/s10773-010-0382-6.

\bibitem{AQKD_Li_2007}
C.-M. Li, T.~Hwang, and K.-C. Lee, ``Security of ''asymmetrical quantum key
  distribution protocol'','' {\em International Journal of Modern Physics C},
  vol.~18, pp.~157--161, 2007.

\bibitem{MQKD_Tsai_2010}
C.-W. Tsai, S.-H. Wang, and T.~Hwang, ``Comment on "n quantum channel are
  sufficient for multi-user quantum key distribution protocol between n
  users",'' {\em Optics Communications}, vol.~283, no.~24, pp.~5285 -- 5286,
  2010.

\bibitem{QKD_Hwang_2011}
T.~Hwang, C.-C. Hwang, and C.-W. Tsai, ``Quantum key distribution protocol
  using dense coding of three-qubit w state,'' {\em The European Physical
  Journal D - Atomic, Molecular, Optical and Plasma Physics}, vol.~61,
  pp.~785--790, 2011.
\newblock 10.1140/epjd/e2010-10320-y.

\bibitem{QKD_Lee_2007}
T.~Hwang, K.-C. Lee, and C.-M. Li, ``Provably secure three-party authenticated
  quantum key distribution protocols,'' {\em IEEE Transactions on Dependable
  and Secure Computing}, vol.~4, no.~1, pp.~71 -- 80, 2007.

\bibitem{QKD_QA_Wei_2011}
T.-S. Wei, C.-W. Tsai, and T.~Hwang, ``Comment on "quantum key distribution and
  quantum authentication based on entangled state",'' {\em International
  Journal of Theoretical Physics}, pp.~1--5, 2011.
\newblock 10.1007/s10773-011-0768-0.

\bibitem{QKD_Shih_2009}
H.-C. Shih, K.-C. Lee, and T.~Hwang, ``New efficient three-party quantum key
  distribution protocols,'' {\em IEEE Journal of Selected Topics in Quantum
  Electronics}, vol.~15, no.~6, pp.~1602 -- 1606, 2009.

\bibitem{MQSS_Hwang_2011}
T.~Hwang, C.-C. Hwang, and C.-M. Li, ``Multiparty quantum secret sharing based
  on ghz states,'' {\em Physica Scripta}, vol.~83, no.~4, p.~045004, 2011.

\bibitem{MQSS_Li_2006}
C.-M. Li, C.-C. Chang, and T.~Hwang, ``Comment on "quantum secret sharing
  between multiparty and multiparty without entanglement'','' {\em Phys. Rev.
  A}, vol.~73, p.~016301, Jan 2006.

\bibitem{MQSS_Lin_2011}
J.~Lin and T.~Hwang, ``An enhancement on shi et al.'s multiparty quantum secret
  sharing protocol,'' {\em Optics Communications}, vol.~284, no.~5, pp.~1468 --
  1471, 2011.

\bibitem{MQSS_Shi_2010}
R.-H. Shi, L.-S. Huang, W.~Yang, and H.~Zhong, ``Quantum secret sharing between
  multiparty and multiparty with bell states and bell measurements,'' {\em
  SCIENCE CHINA Physics, Mechanics and Astronomy}, vol.~53, pp.~2238--2244,
  2010.
\newblock 10.1007/s11433-010-4181-0.

\bibitem{MQSS_Wang_2010}
S.-H. Wang, S.-K. Chong, and T.~Hwang, ``On "multiparty quantum secret sharing
  with bell states and bell measurements",'' {\em Optics Communications},
  vol.~283, no.~21, pp.~4405 -- 4407, 2010.

\bibitem{MQSS_Yan_2005}
F.-L. Yan and T.~Gao, ``Quantum secret sharing between multiparty and
  multiparty without entanglement,'' {\em Phys. Rev. A}, vol.~72, p.~012304,
  Jul 2005.

\bibitem{QSS_Chen_2009}
J.-H. Chen, K.-C. Lee, and T.~Hwang, ``The enhancement of zhou et al.'s quantum
  secret sharing protocol,'' {\em International Journal of Modern Physics C},
  vol.~20, pp.~1531--1535, 2009.

\bibitem{QSS_Gottesman_2000}
D.~Gottesman, ``Theory of quantum secret sharing,'' {\em Phys. Rev. A},
  vol.~61, p.~042311, Mar 2000.

\bibitem{QSS_Hsieh_2010}
C.-R. Hsieh, C.-W. Tasi, and T.~Hwang, ``Quantum secret sharing using ghz-like
  state,'' {\em Communications in Theoretical Physics}, vol.~54, no.~6,
  p.~1019, 2010.

\bibitem{QSS_QECC_Zhang_2011}
Z.-R. Zhang, W.-T. Liu, and C.-Z. Li, ``Quantum secret sharing based on quantum
  error-correcting codes,'' {\em Chinese Physics B}, vol.~20, no.~5, p.~050309,
  2011.

\bibitem{QRC_Huelga_2001}
S.~F. Huelga, J.~A. Vaccaro, A.~Chefles, and M.~B. Plenio, ``Quantum remote
  control: Teleportation of unitary operations,'' {\em Phys. Rev. A}, vol.~63,
  p.~042303, Mar 2001.

\bibitem{QRC_Chen_2005}
L.-B. Chen, H.~Lu, and Y.-H. Liu, ``Implementing remotely a single-qubit
  rotation operation by three-qubit entanglement,'' {\em Chinese Physics},
  vol.~14, no.~7, p.~1323, 2005.

\bibitem{QRC_Chen_2007}
A.-X. Chen, L.~Deng, and Q.-P. Wu, ``Remote operation on quantum state among
  multiparty,'' {\em Communications in Theoretical Physics}, vol.~48, no.~5,
  p.~837, 2007.

\bibitem{QRC_Huelga_2002}
S.~F. Huelga, M.~B. Plenio, and J.~A. Vaccaro, ``Remote control of restricted
  sets of operations: Teleportation of angles,'' {\em Phys. Rev. A}, vol.~65,
  p.~042316, Apr 2002.

\bibitem{QRC_Yao_2006}
C.-M. Yao, ``Quantum remote control of unitary operations on a qubit of pure
  entangled states,'' {\em Chinese Physics Letters}, vol.~23, no.~3, p.~545,
  2006.

\bibitem{QRC_Zhang_2004}
Y.-Z. Zheng, P.~Ye, and G.-C. Guo, ``Probabilistic implementation of non-local
  cnot operation and entanglement purification,'' {\em Chinese Physics
  Letters}, vol.~21, no.~1, p.~9, 2004.

\end{thebibliography}

\end{document}